\documentclass[twocolumn,showpacs,aps,pre]{revtex4-1} 
\usepackage{amsmath,amssymb}
\usepackage{graphicx}
\usepackage{color}
\usepackage{xcolor}
\usepackage{stmaryrd}
\usepackage{wasysym}
\usepackage[utf8]{inputenc}

\newcommand{\com}[1]{{#1}}

\begin{document}

\title{Scaling of far-field wake angle of non-axisymmetric pressure disturbance}
\author{F. Moisy}
\author{M. Rabaud}
\affiliation{Universit\'e Paris-Sud, CNRS, Laboratoire FAST. B\^atiment 502, 91405 Orsay, France.}

\date{\today}

\begin{abstract}
It has been recently emphasized that the angle of maximum wave amplitude $\alpha$ in the wake of a disturbance of finite size can be significantly narrower than the maximum value $\alpha_K = \sin^{-1}(1/3) \simeq 19.47^\mathrm{o}$ predicted by the classical analysis of Kelvin. For axisymmetric disturbance, simple argument based on the Cauchy-Poisson initial-value problem suggests that the wake angle decreases following a Mach-like law at large velocity, $\alpha \simeq Fr_L^{-1}$ , where $Fr_L=U/\sqrt{gL}$ is the Froude number based on the disturbance velocity $U$, its size $L$, and gravity $g$. In this paper we extend this analysis to the case of non-axisymmetric disturbances, relevant to real ships. We find that, for intermediate Froude numbers, the wake angle follows an intermediate scaling law $\alpha \simeq Fr_L^{-2}$, in agreement with the recent prediction of Noblesse \textit{et al.} [Eur. J. Mech. B/Fluids {\bf 46}, 164 (2014)]. We show that beyond a critical Froude number, which scales as $A^{1/2}$ (where $A$ is the length-to-width aspect ratio of the disturbance), the asymptotic scaling $\alpha \simeq Fr_B^{-1}$ holds, where now $Fr_B = A^{1/2} Fr_L$ is the Froude number based on the disturbance width. We propose a simple model for this transition, and provide a regime diagram of the scaling of the wake angle as a function of parameters $(A,Fr_L)$.
\end{abstract}

\maketitle

\section{Introduction}
\label{sec_intro}

Lord Kelvin was the first to explain why a ship moving at constant velocity in deep water generates waves confined in a triangular wedge~\cite{Kelvin1887,Darrigol}. He demonstrated that the stationary wave pattern is composed of a transverse and a divergent wave system delimited by a cusp line making a constant half-angle $\alpha_K=\sin^{-1} (1/3) \simeq 19.47^\mathrm{o}$ with the ship trajectory~\cite{Havelock1908,Lamb,Ursell1960,Wehausen,Lighthill}. \com{Recently, airborne images of ship wakes showing angle of maximum wave amplitude significantly smaller than the Kelvin prediction have been analyzed~\cite{Rabaud2013}, renewing the interest in this classical subject~\cite{Darmon2014,Ellingsen2014,Noblesse2014,Benzaquen2014,Moisy2014}. We propose here to extend the phenomenological approach introduced in Ref.~\cite{Rabaud2013} to non-axisymmetric disturbances, providing a more realistic description of elongated boats.}

For pure gravity waves in deep water, the governing parameter for the wake angle is the Froude number based on the hull length $L$, $Fr_L=U/\sqrt{gL}$, which is the ratio of the boat velocity $U$ and velocity of gravity waves of wavelength of the order of $L$~\cite{Wehausen,Lighthill}. Ignoring the exact shape of the boat and retaining $L$ as the unique length scale of the problem, it is possible to infer the scaling of the wake angle from the following general property of dispersive waves: A disturbance of size $L$ mostly excites a wave packet containing wave numbers $k_f$ of order $L^{-1}$ and propagating at the group velocity $c_g  = \frac{1}{2} \sqrt{g/k_f}$. This is the main result of the Cauchy-Poisson initial-value problem, first analyzed in 1815 \cite{Darrigol,Havelock1908,Lamb,Wehausen,Lighthill}. It follows that the energy emitted by a disturbance of finite size is effectively radiated at a constant group velocity. Accordingly, the maximum amplitude of the waves at large $Fr_L$ is found at the Mach-like angle $\alpha \simeq c_g / U \simeq Fr_L^{-1}$ (this law does not apply for moderate $Fr_L$ because of the cusp in the wave pattern, which concentrates the maximum amplitude at the Kelvin angle $\alpha_K$). This law is compatible with the airborne images of ship wakes and numerical simulations of Ref.~\cite{Rabaud2013}, and has recently received mathematical confirmation by Darmon {\it et al.} \cite{Darmon2014} for an axisymmetric disturbance.


Recently Noblesse \textit{et al.} \cite{Noblesse2014} proposed an alternate scaling for the decrease of the wake angle at large velocity, $\alpha \simeq Fr_L^{-2}$, which turns out to also fit well the airborne data of Ref.~\cite{Rabaud2013}, at least in an intermediate range of Froude numbers. Their analysis relies on the modeling of a real boat as two out-of-phase point sources separated by a distance of order $L$. This simple model classically reproduces the double Kelvin wedge originating at the bow and the stern of poorly streamlined boats at small Froude numbers \cite{Lamb}. In their approach the decrease of the wake angle is described in terms of destructive interferences between the two Kelvin patterns which occur when the wavelength of the transverse waves becomes of the order or larger than the hull length.

\com{The aim of this paper is to investigate the influence of the aspect ratio of a non-axisymmetric disturbance on the far-field angle of maximum wave amplitude, focusing on the case of pure gravity linear waves in deep water.  We consider the simplest non-axisymmetric disturbance, a Gaussian pressure field of elliptic iso-level lines with aspect ratio $A = L/B$, where $L$ and $B$ are the disturbance length and width ({\it beam}), respectively.  Of course this crude simplification does not pretend to reproduce the complexity of real ship wakes. Hulls are rigid objects that cannot be reduced to a simple applied pressure disturbance, with a trim being moreover a function of the ship velocity (the effective aspect ratio is a decreasing function of the Froude number for rapid boats in the planing regime). An asymptotic analysis of the Fourier integral defining the surface elevation in this linear model can be performed to compute the wake angle as a function of the Froude number and the disturbance aspect ratio~\cite{Benzaquen2014}. Here we propose a simple geometrical approach to determine the scaling of this wake angle, by extending the phenomenological model of Ref.~\cite{Rabaud2013,Moisy2014} to non-axisymmetric disturbances.} A regime diagram in terms of these two parameters is proposed, which brings together the axisymmetric disturbance regime ($\alpha \simeq Fr_B^{-1}$ for $A=1$) of Refs. \cite{Rabaud2013, Darmon2014, Benzaquen2014} and the strongly elongated disturbance regime ($\alpha \simeq Fr_L^{-2}$ for $A \gg 1$) of Ref. \cite{Noblesse2014}.

\begin{figure}
\centerline{\includegraphics[width=0.78\linewidth]{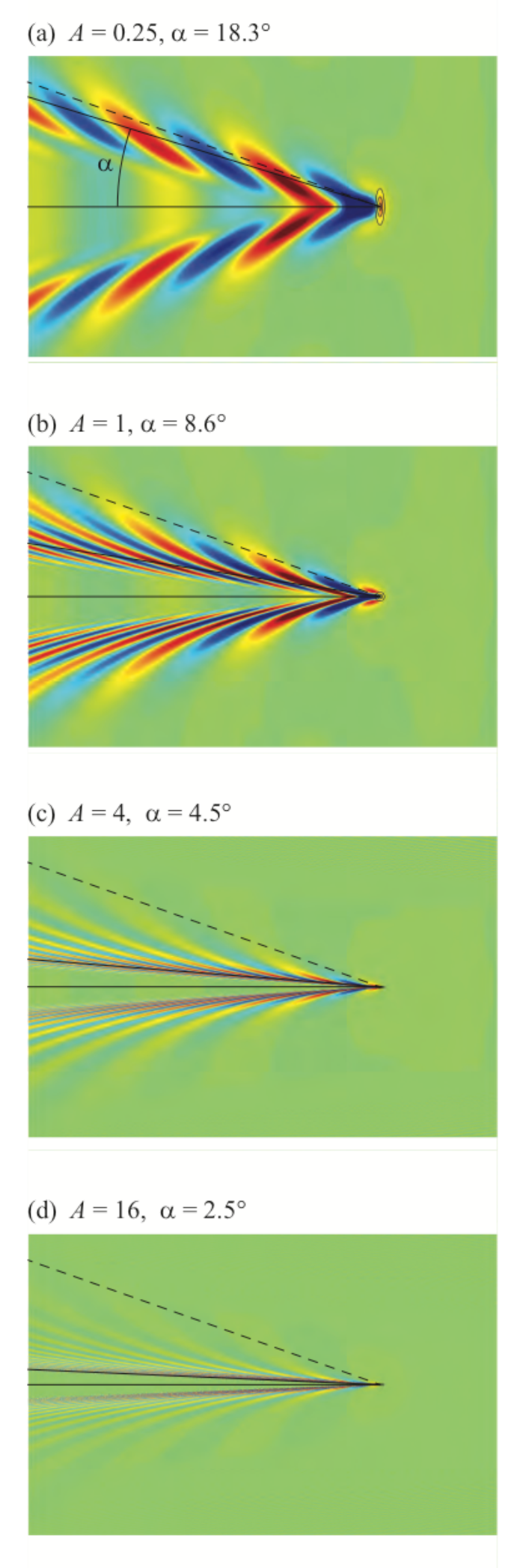}}
\caption{(Color online) Wave field computed for the non-axisymmetric Gaussian pressure disturbance (\ref{eq:pressure_ani}) at fixed longitudinal Froude number $Fr_L= 1.5$ for various aspect ratio: (a) $A=0.25$,  (b) $A=1$, (c) $A=4$, (d) $A=16$. Color maps and scales are the same for the four images. Dashed line: classical Kelvin angle $\alpha_K$; solid line: angle $\alpha$ of maximum wave amplitude.}
\label{fig:Wakes_FrL}
\end{figure}

\section{Numerical simulations}
\label{sec:simulation}

The wave pattern is computed using the classical simplification due to Havelock \cite{Havelock1908}, in which the motion of a rigid hull is modeled by the translation at the water surface of a pressure disturbance $P(\textbf{r})$.
The resulting wave field is given by \cite{Darmon2014,Lighthill,Raphael1996}: 
\begin{equation}
\zeta({\bf x}) = - \lim_{\epsilon \rightarrow 0} \frac{1}{(2 \pi)^2} \int \!\!\!\! \int \frac{k \hat P ({\bf k}) / \rho}{\omega({\bf k})^2 - ({\bf k} \cdot {\bf U} - i \epsilon)^2} e^{i {\bf k} \cdot {\bf x}} d^2 {\bf k},
\label{eq:zetint}
\end{equation}
with $\hat P({\bf k}) $ the two-dimensional Fourier transform of $P(\textbf{r})$, $\rho$ the fluid density, ${\bf U} = U {\bf e}_x$ the disturbance velocity, $\omega({\bf k}) = \sqrt{g |{\bf k}|}$ the wave frequency for pure gravity waves in deep water, and $\epsilon>0$ a small parameter introduced to avoid the divergence of the integrand.
In Refs.~\cite{Rabaud2013,Darmon2014,Ellingsen2014,Moisy2014} an axisymmetric  pressure distribution is used. Here we use a Gaussian pressure distribution with elliptical iso-values of longitudinal axis $L$ (along the disturbance motion ${\bf e}_x$) and transverse axis $B$,
\begin{equation}
P({\bf r}) = P_0 \exp \left[- \pi^2 \left( \frac{x^2}{L^2} + \frac{y^2}{B^2} \right) \right].
\label{eq:pressure_ani}
\end{equation}
For pure gravity waves in deep water, there are three length scales in the problem, $L$, $B$ and $U^2/g$, so the far-field wake angle of this non-axisymmetric disturbance is governed by two independent non-dimensional parameters.
The first one is the aspect ratio
\begin{equation}
A=L/B.
\label{eq:aspect_ratio}
\end{equation}
The second one can be either the {\it longitudinal} Froude number based on the disturbance length $L$,
$$
Fr_L=\frac{U}{\sqrt{gL}}
$$
or the {\it transverse} Froude number based on the disturbance width $B$,
$$
Fr_B = \frac{U}{\sqrt{gB}} = A^{1/2} Fr_L.
$$
Both sets of non-dimensional numbers ($A,Fr_L$) and ($A,Fr_B$) turn out to be useful to describe the various wake angle regimes in the following.

We have computed the wake pattern for a wide range of aspect ratios, from $A=0.25$ (ellipse traveling along its smallest dimension) to $A=64$ (very thin ellipse traveling along its largest dimension), and for Froude numbers $Fr_L$ ranging from 0.1 to 100. These numbers go well beyond realistic values for ships \com{(typically $Fr_L \simeq 0.1-2$ and $A \simeq 2-10$)}, but they are nonetheless useful to infer asymptotic scaling laws for the wake angle. The Fourier integral (\ref{eq:zetint}) is integrated on a square domain of size $L_{box}$, discretized on a grid of $N \times N$ collocation points.  For an aspect ratio $A \gg 1$, the resolution $N$ must be such that $L_{box} \gg L \gg B \gg L_{box}/N$. We take here $N=3 \times 2^{12} = 12288$, which is sufficient to simulate  the wake pattern for the largest aspect ratio, $A=64$.

\begin{figure}[hbt]
\centerline{\includegraphics[width=0.95\linewidth]{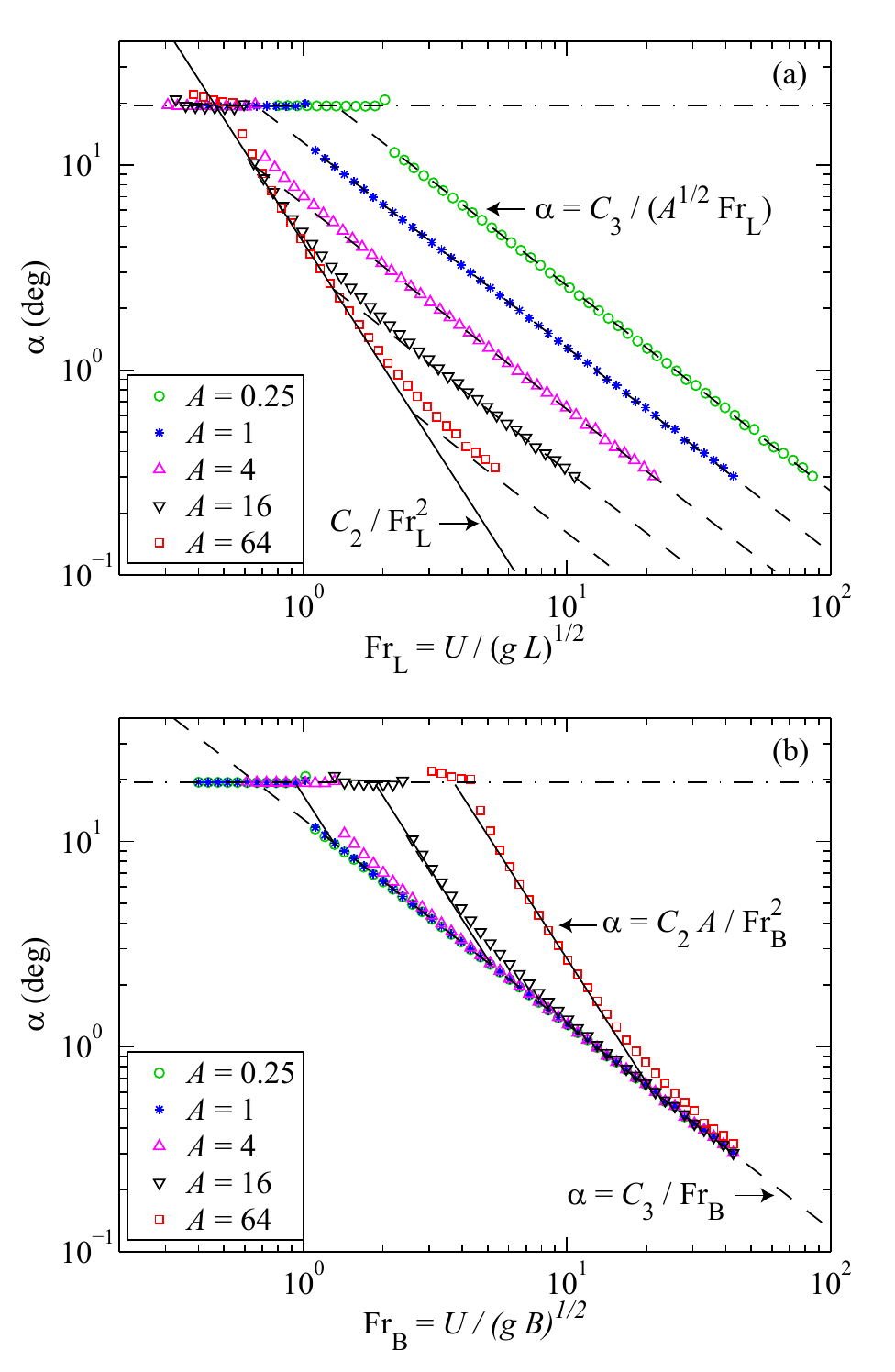}}
\caption{(Color online) Angle of maximum wave amplitude $\alpha$ as a function of the longitudinal Froude number $Fr_L$ (a) and the transverse Froude number $Fr_B$ (b), for a non-axisymmetric Gaussian pressure disturbance at five values of the aspect ratio, $A=L/B=0.25, 1, 4, 16$ and 64.}
\label{fig:Fr_Long}
\end{figure}

Typical  wave patterns are shown in Fig. \ref{fig:Wakes_FrL} for a constant longitudinal Froude number $Fr_L = 1.5$ with varying aspect ratios $A$ from 0.25 to 16. The far-field wake angle $\alpha$ is measured as the angle between the disturbance trajectory and the line going through the maximum amplitude of the waves. For this value of $Fr_L$ the wake angle $\alpha$ is smaller than the Kelvin value, and is clearly a decreasing function of the aspect ratio $A$.

Figure \ref{fig:Fr_Long} shows the angle $\alpha$ for various aspect ratio, as a function of both the {\it longitudinal} and {\it transverse} Froude numbers, $Fr_L$ (Fig.~\ref{fig:Fr_Long}a) and $Fr_B$ (Fig.~\ref{fig:Fr_Long}b). From these two sets of non-dimensional parameters it is possible to identify the following three regimes:

(1) At low velocity, angles $\alpha$ close to the classical Kelvin angle $\alpha_K \simeq 19.47^\mathrm{o}$ are found. This regime is valid up to $Fr_{L} \simeq 0.5$ for $A > 1$ [see Fig.~\ref{fig:Fr_Long}(a)], whereas it is valid up to $Fr_{B} \simeq 0.5$ for $A < 1$, i.e. up to $Fr_{L} \simeq 0.5 A^{-1/2}$ [see Fig.~\ref{fig:Fr_Long}(b)]. In other words, the Kelvin regime holds when the Froude number based on the smallest size of the disturbance is below 0.5.

(2) At intermediate velocity, provided that the aspect ratio $A$ is sufficiently large, the wake angle is governed by the {\it longitudinal} Froude number, and follows the law
\begin{equation}
\alpha \simeq \frac{C_2}{Fr_L^2}
\label{eq:Froude_W}
\end{equation}
with $C_2 \approx 0.073 \pm 0.003$. This intermediate regime is compatible with the analysis of Noblesse {\it et al.}~\cite{Noblesse2014}. It must be noted that its extent is moderate: For $A=64$ (a value unrealistically large for real ships), this scaling holds in the range $0.5 < Fr_L < 2$ only.

(3) At larger velocity, the wake angle is governed now by the {\it transverse} Froude number $Fr_B$, and follows the law
\begin{equation}
\alpha \simeq \frac{C_3}{Fr_B}
\label{eq:Froude_L}
\end{equation}
with $C_3 \approx 0.22 \pm 0.01$. This law is in excellent agreement with the analytical prediction $C_3 = 1/(\pi^{1/2} 40^{1/4}) \simeq 0.224$ of Ref.~\cite{Darmon2014} for an axisymmetric Gaussian pressure disturbance [Eq.~(\ref{eq:pressure_ani}) with $A=1$].

\begin{figure}[hbt]
\begin{center}
\includegraphics[width=0.9\linewidth]{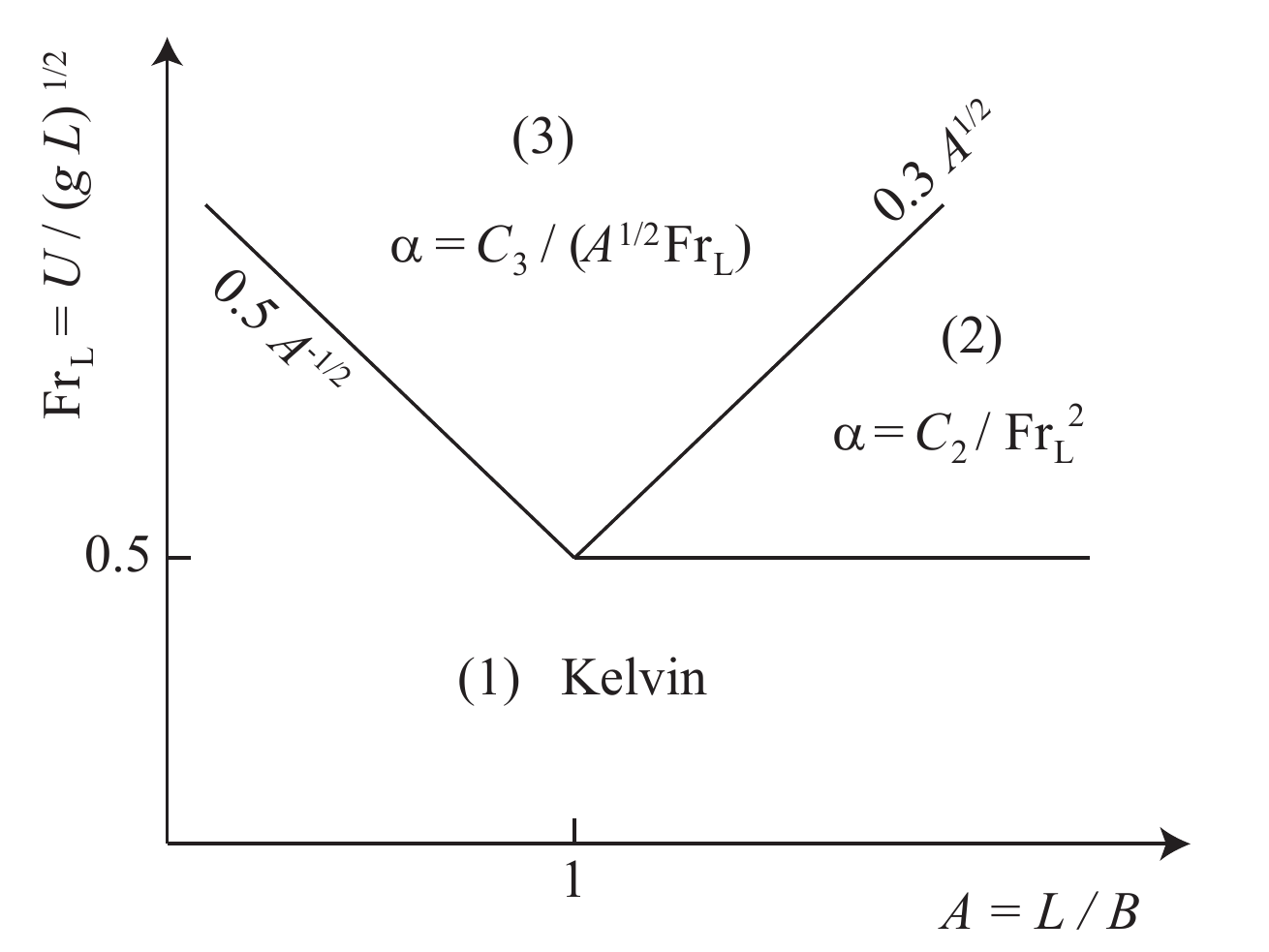}
\caption{The three regimes of wake angle in the plan of parameters $(A, Fr_L)$ in logarithmic scales.}
\label{fig:diag}
\end{center}
\end{figure}

The three wake regimes are summarized in the plan of parameters $(A, Fr_L)$ in Fig.~\ref{fig:diag}.
The boundary between regimes (2) and (3) is given by
$Fr_L \simeq (C_2/C_3) A^{1/2} \simeq 0.33 A^{1/2}$.
Interestingly, for nearly axisymmetric disturbances (typically $A < 2$), only the first and third regimes are observed: the wake angle directly proceeds from $\alpha \simeq 19.47^\mathrm{o}$  to $\alpha \simeq Fr_B^{-1}$ as the velocity is increased, in agreement with the analysis of Refs. \cite{Rabaud2013,Darmon2014} for axisymmetric disturbance.

\section{Phenomenological model}
\label{sec:modelisation}

We introduce in the following a simple model which describes the transition between the three wake regimes found in the simulation. The basic assumption is that the amplitude of the waves excited by a moving disturbance is small when their wave lengths are much larger or much smaller than the disturbance size. This hypothesis is a direct consequence of the Cauchy-Poisson initial value problem. It was first used in Ref.~\cite{Rabaud2013} for axisymmetric disturbance, and we extend this approach here to the case of non-axisymmetric disturbance.

The physical content of the Cauchy-Poisson problem can be described as follows.
When a stone of size $L$ is thrown in a pound, a wave packet containing all the wavelengths is excited. However, in the far field, waves of significant amplitude have their wave length of the order of the stone size. Let us consider for simplicity the case of a Gaussian initial perturbation of the interface, which excites a Gaussian spectrum of wave numbers, each propagating with its group velocity $c_g(k)=\frac 12 \sqrt{g/k}$. In this spectrum, the small wave numbers of large initial amplitude propagate faster, so they spread over a larger distance and their spatial density of energy (and hence their amplitude) rapidly decreases. On the other hand, the large wave numbers do not spread much, but they are of small initial amplitude. As a result, the waves of intermediate wave number $k$, of the order of $L^{-1}$, are of maximum amplitude.

\begin{figure}[hbt]
\centerline{\includegraphics[width=0.6\linewidth]{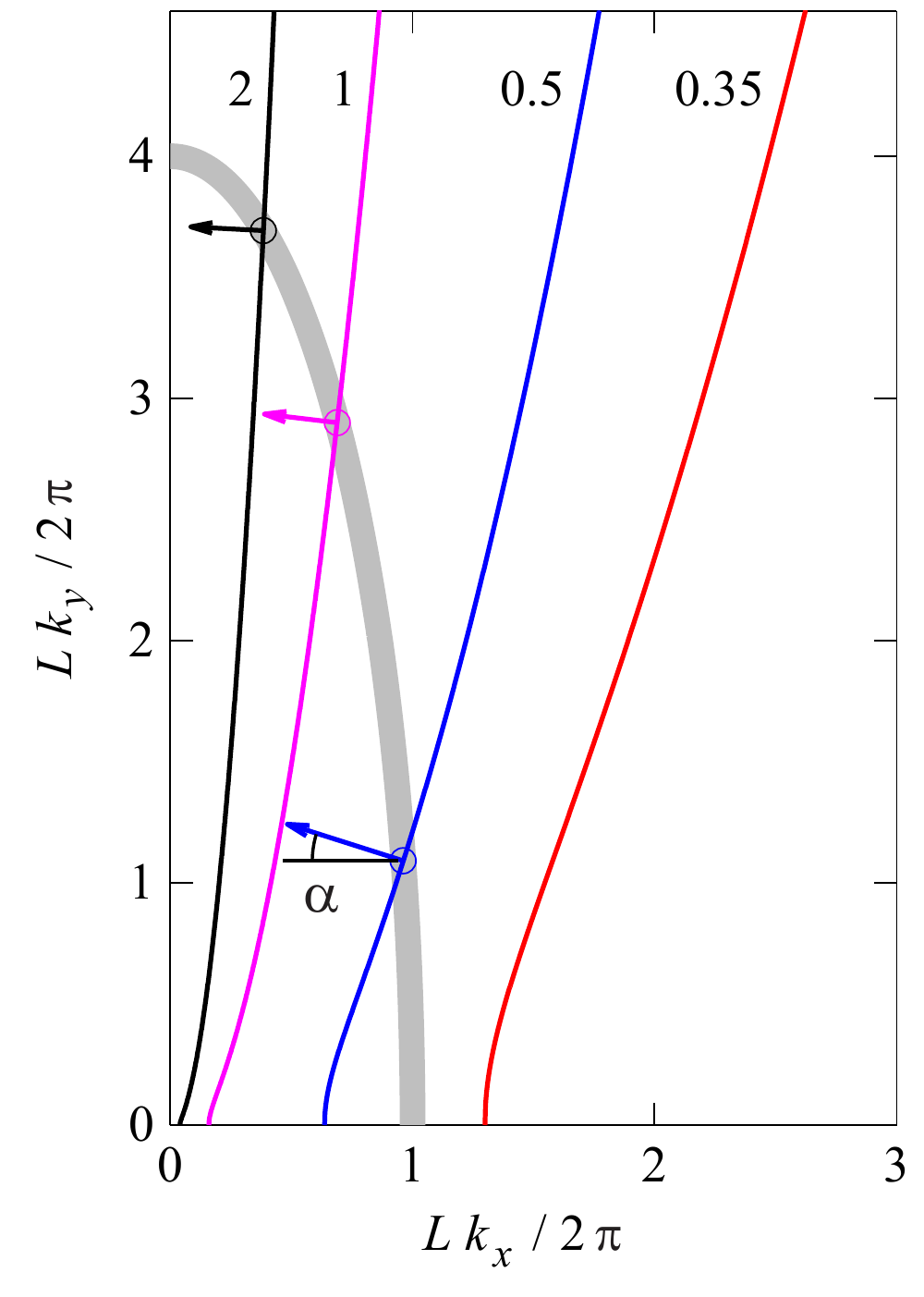}}
\caption{(Color online) Graphical solution in the Fourier space for the wake angle of a non-axisymmetric disturbance. Thin (color) lines: Wave vectors ${\bf k}$ satisfying the stationary condition $\Omega({\bf k}) = 0$ [Eq.~(\ref{eq:loc})], for longitudinal Froude numbers $Fr_L$ = 0.35, 0.5, 1 and 2. The normal to the curve gives the group velocity relative to the disturbance, ${\bf c}'_g$, making angle $\alpha(k)$ with the axis $x$. Thick (gray) line: Energy-containing domain radiated by a disturbance [Eq.~(\ref{eq:ellipse}), shown here for $A=4$]. The wake angle $\alpha$ for a disturbance of a given Froude number is given by the normal to the curve $\Omega({\bf k}) = 0$ taken at the point of intersection $(\tilde k_x, \tilde k_y)$ between the bold gray line and the thin line.}
\label{fig:locus_aniso}
\end{figure}

We start by describing the wake pattern in the Fourier space. The waves in the wake being stationary in the frame of reference of the disturbance traveling at velocity ${\bf U} = U {\bf e}_x$, their wave vectors must be such that their relative (i.e., Doppler-shifted) frequency, $\Omega({\bf k}) = \omega({\bf k}) - {\bf k} \cdot {\bf U}$, is zero  [$\omega({\bf k}) = \sqrt{g k}$ is the dispersion relation of gravity waves, with $k = |{\bf k}|$]. Introducing the non-dimensional wave vector ${\bf K} = {\bf k} / k_g$, with $k_g = g/U^2$, the stationary condition $\Omega({\bf k}) = 0$ writes
\begin{equation}
K_y^2 = K_x^2 (K_x^2-1).
\label{eq:loc}
\end{equation}
This relation is plotted in Fig. \ref{fig:locus_aniso} (it corresponds to a family of curves because of the use of the axis normalized by the disturbance size---see below). The energy of a given wave vector ${\bf K}$ propagates according to its relative group velocity (in the frame of the disturbance), ${\bf c}'_g = \nabla_k \Omega = {\bf c}_g - {\bf U}$, where ${\bf c}_g
= \nabla_k \omega$ is the group velocity in the frame of the liquid at rest. Accordingly, the relative group velocity ${\bf c}'_g$ is a vector normal to the curves (\ref{eq:loc}), as illustrated by the arrows in Fig.~\ref{fig:locus_aniso}~\cite{Rousseaux2013,Doyle2013}. The angle of ${\bf c}'_g$ with respect to $- {\bf U}$, which we call the {\it radiation angle}, is therefore given by
\begin{equation}
\tan \alpha(K) = \left( \frac{\partial K_y}{\partial K_x} \right)^{-1} = \frac{\sqrt{K_x^2 - 1}}{2 K_x^2 - 1}
\label{eq:tana}
\end{equation}
(see Ref. \cite{Rabaud2013,Moisy2014} for an alternate derivation of this angle in the physical space). This radiation angle is 0 for $K_x = 1$ and for $K_x \rightarrow \infty$, and reaches the maximum $\alpha_K = \tan^{-1} (1/\sqrt{8}) \simeq 19.47^\mathrm{o}$ \com{at the inflection point of the curve (\ref{eq:loc}), which is located at ${\bf K}_0 = (\sqrt{3/2}, \sqrt{3/4})$, with $K_0 = |{\bf K_0}| = 3/2$.}

For a disturbance characterized by a given spectrum, the energy density of each wave number of the spectrum propagates in the direction given by Eq.~(\ref{eq:tana}). The resulting {\it wake angle} $\alpha$, i.e. the angle of maximum wave amplitude, is therefore given by the radiation angle $\alpha(k)$ at which most of the energy supplied by the disturbance is effectively radiated. Two cases must be considered:
(i) if the maximum of the radiated energy is in the vicinity of the inflection point
$K_0 = 3/2$, the radiated energy focuses along the Kelvin angle $\alpha_K$;
(ii) otherwise, the wake angle is given by the radiation angle evaluated at the most excited wave number.

In order to determine the wake angle it is now necessary to model the effect of the non-axisymmetric disturbance in the Fourier space. We extend the analysis stemming from the Cauchy-Poisson problem by assuming that the wave of maximum amplitude in each direction has a wavelength of the order of the disturbance size along that direction. Accordingly, the energy-containing domain in the spectral space is an ellipse, of semi-axes $2\pi/L$ and $2\pi/B = 2\pi A/L$:
\begin{equation}
\left(\frac{L k_x}{2 \pi}\right)^2 + \left(\frac{L k_y}{2 \pi A}\right)^2 = 1
\label{eq:ellipse}
\end{equation}
(plotted as the bold gray line in Fig. \ref{fig:locus_aniso} in the case $A=4$).
\com{In practice, because of the finite extent of the disturbance, this energy-containing ellipse has a thickness of the order of unity in the spectral space; for simplicity we do not consider this thickness in the following.}
Normalizing Eq.~(\ref{eq:ellipse}) by $k_g = g/U^2$ and using the relation $L k_g = Fr_L^{-2}$, the ellipse  writes
\begin{equation}
K_x^2 + (K_y / A)^2 = (2 \pi)^2 Fr_L^4.
\label{eq:dist}
\end{equation}

For a given Froude number $Fr_L > 1/\sqrt{2\pi} \simeq 0.4$, the energy-containing ellipse (\ref{eq:dist}) intersects the stationary curve $\Omega({\bf k})=0$ (\ref{eq:loc}) at the point $\tilde{\bf K} = (\tilde K_x, \tilde K_y)$ satisfying
\begin{equation}
\tilde{K}_x^2 = \frac{1}{2} \left[ 1-A^2 \pm \sqrt{(A^2-1)^2 + 4 A^2 (2 \pi)^2 Fr_L^4} \right]
\label{eq:kxt}
\end{equation}
(where only the sign $+$ has a physical meaning). If this intersection falls in the vicinity of the inflection point, $\tilde{\bf K} \simeq {\bf K}_0$ [case (i) above], the energy radiated by the disturbance focuses at the Kelvin angle. Otherwise, energy is mostly found at the radiation angle (\ref{eq:tana}) evaluated at the intersection point (\ref{eq:kxt}). \com{This geometrical construction can be readily generalized for any dispersion relation, e.g. for finite-depth gravity waves, for capillary-gravity waves, etc.}

\begin{figure}[hbt]
\centerline{\includegraphics[width=0.9\linewidth]{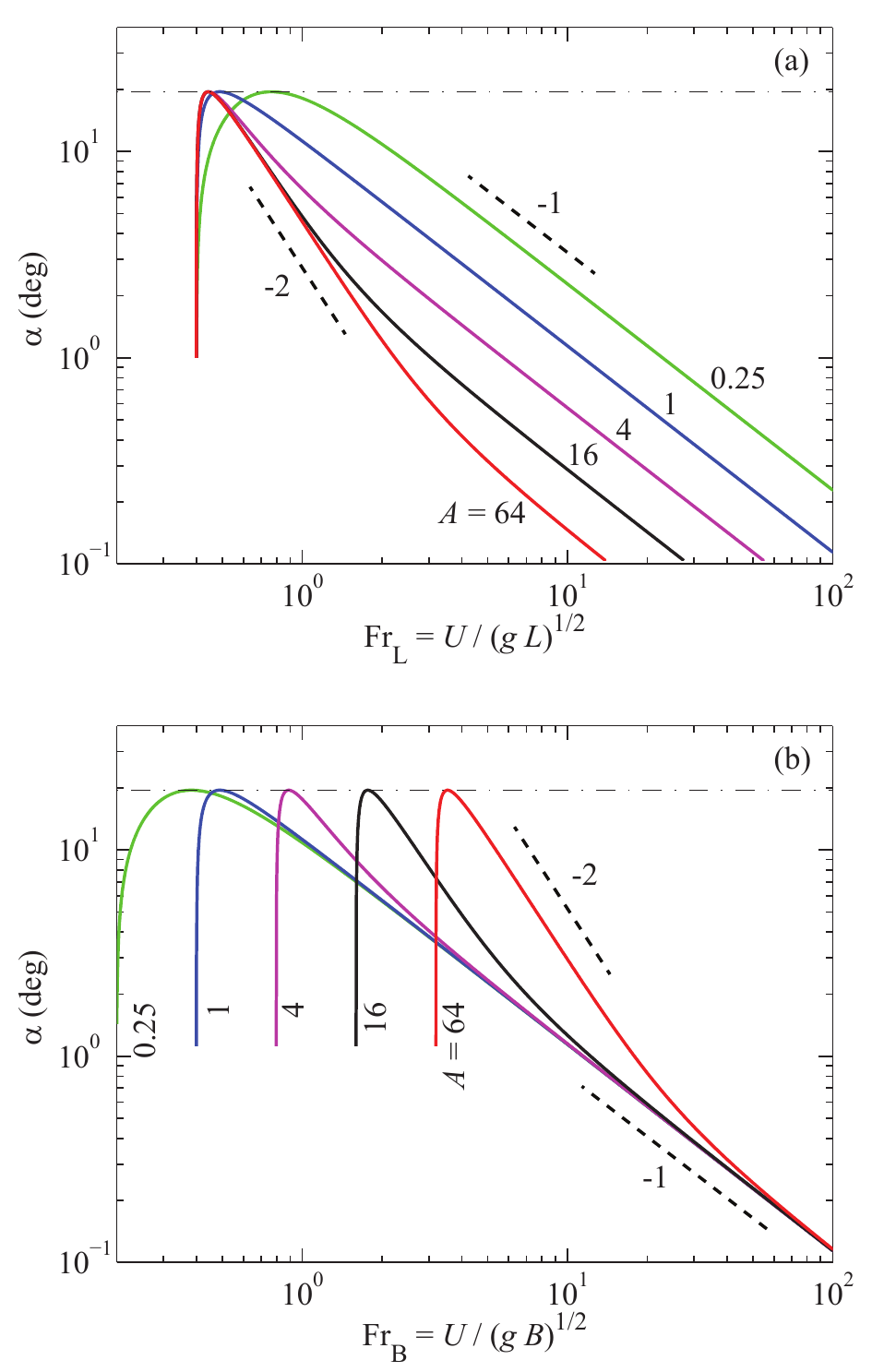}}
\caption{(Color online) Wake angle according to the model (\ref{eq:tana})-(\ref{eq:kxt}), as a function of the longitudinal Froude number $Fr_L$ (a) and transverse Froude number $Fr_B$ (b), for aspect ratios $A$ = 0.25, 1, 4, 16 and 64. \label{fig:angle_vs_fr}}
\end{figure}

The wake angle $\alpha$ given by this model (\ref{eq:tana})-(\ref{eq:kxt}) is plotted as a function of $Fr_L$ and $Fr_B$ for various aspect ratios $A$ in Fig.~\ref{fig:angle_vs_fr}. For parameters such that the wake angle is a decreasing function of $Fr_L$, the model shows an excellent agreement with the wake angle determined numerically in Fig.~\ref{fig:Fr_Long} for the elliptical Gaussian disturbance. 
On the other hand, since the model does not contain the physics of the wave focusing along the inflection point (cusp angle), it cannot describe the wake angle
at small Froude numbers, when $\alpha$ is close to the Kelvin angle, and the sharp jump to smaller angles observed in Fig.~\ref{fig:Fr_Long} as the Froude number is increased.

The two scaling laws found numerically [Eq.~(\ref{eq:Froude_W}) and (\ref{eq:Froude_L})] are readily recovered in the case of a very elongated disturbance ($A \gg 1$):

For $1 \ll Fr_L \ll \sqrt{A}$, Eq.~(\ref{eq:kxt}) reduces to $\tilde K_x \simeq 2\pi Fr_L^2$, yielding the wake angle:
$$
\alpha \simeq \frac{1}{4\pi Fr_L^2}.
$$
This is the intermediate regime (2) of Fig.~\ref{fig:diag}, in which the wake angle does not depend on the aspect ratio $A$, and is governed by $Fr_L$ only. The numerical constant $C_2 = 1/4\pi \simeq 0.080$ is very close to the one found numerically, $C_2 \simeq 0.073 \pm 0.003$.

For $Fr_L \gg \sqrt{A}$, Eq.~(\ref{eq:kxt}) reduces to $\tilde K_x \simeq (2\pi A)^{1/2} Fr_L$, yielding the wake angle:
$$
\alpha \simeq \frac{1}{2 \sqrt{2 \pi} \sqrt{A} Fr_L} = \frac{1}{2 \sqrt{2 \pi} Fr_B}.
$$
This is the asymptotic regime (3) of Fig.~\ref{fig:diag}, in which the wake angle is governed now by the transverse Froude number  $Fr_B = U/\sqrt{gB}$.
The numerical constant is $C_3 = 1/(2 \sqrt{2\pi}) \simeq 0.20$, very close to the one found numerically, $C_3 \simeq 0.22 \pm 0.01$.

\section{Conclusion}
\label{sec:conclusion}

We have investigated, by means of numerical simulations and a simple phenomenological model, the influence of the aspect ratio $A$ of the disturbance on the scaling of the angle of maximum wave amplitude. The model stems from the general property of dispersive waves that a disturbance of finite size excites a wave packet containing wave lengths of order of the disturbance size, which we apply here to the case of non-axisymmetric disturbances. In spite of its simplicity,
the present model successfully reproduces the two wake regimes reported in the literature, governed by the Froude number based either on the disturbance length $L$ or width $B$: For axisymmetric or weakly elongated disturbances the asymptotic law $\alpha \simeq Fr_B^{-1}$ of Refs. \cite{Rabaud2013, Darmon2014} is recovered, whereas for elongated disturbances an intermediate scaling $Fr_L^{-2}$ is found. This intermediate scaling is compatible with the analysis of  Noblesse {\it et al.}~\cite{Noblesse2014}, which applies for two separated point sources.

Of course the application of this highly simplified model for real ship wakes is questionable, since it ignores the complexity of the flow around real ship hulls (with detached boundary layers, turbulence, wave breaking etc.). In particular, the cross-over between the two scaling laws at $Fr_L \simeq A^{1/2}$, which can be hardly tested from existing data of airborne wake images \cite{Rabaud2013}, should be confirmed by systematic measurements of ship models of various aspect ratios or numerical simulations reproducing realistic hull shapes.

\acknowledgments

We acknowledge M. Benzaquen, A. Darmon, E. Rapha\"el, and G. Rousseaux for fruitful discussions. F.M. acknowledges the Institut Universitaire de France for its support.


\begin{thebibliography}{}

\bibitem{Kelvin1887} Lord Kelvin,
Proc. Inst. Mech. Eng. {\bf 38}, 409 (1887).

\bibitem{Darrigol}	O. Darrigol, \textit{Worlds of Flow: A History of Hydrodynamics from the Bernoullis to Prandtl} (Oxford University Press, U.K., 2005).

\bibitem{Havelock1908} T.H. Havelock,
Proc. R. Soc. London, Ser. A {\bf 81}, 398 (1908).

\bibitem{Lamb} Sir H. Lamb, {\it Hydrodynamics} (Dover, New York, 1945).

\bibitem{Wehausen} J. V. Wehausen and E.V. Laitone,
in Encyclopedia of Physics (Springer, Berlin, 1960), Vol. 9, Part 3, p. 446.

\bibitem{Ursell1960} F. Ursell,
J. Fluid Mech. {\bf 9}, 333 (1960).

\bibitem{Lighthill}  J. Lighthill, {\it Waves in Fluids} (Cambridge University Press, Cambridge, U.K., 1978).


\bibitem{Rabaud2013} M. Rabaud and F. Moisy,
Phys. Rev. Lett. {\bf 110}, 214503 (2013).

\bibitem{Darmon2014} A. Darmon, M. Benzaquen and E. Rapha\"{e}l,
J. Fluid Mech. {\bf 738}, R3 (2014).

\bibitem{Ellingsen2014} S.A. Ellingsen,
J. Fluid Mech. {\bf 742}, R2 (2014).

\bibitem{Noblesse2014}  F. Noblesse, J. He, Y. Zhu, L. Hong, C. Zhang, R. Zhu, and C. Yang,
Eur. J. Mech. B/Fluids {\bf 46}, 164 (2014).

\bibitem{Benzaquen2014} M. Benzaquen, A. Darmon and E. Rapha\"{e}l,
arXiv:1404.1699.

\bibitem{Moisy2014} F. Moisy and M. Rabaud,
arXiv:1406.0422.

\bibitem{Raphael1996} E. Rapha\"{e}l and P.-G. de Gennes, 
Phys. Rev. E {\bf 53} (4), 3448 (1996).

\bibitem{Rousseaux2013} I. Carusotto and G. Rousseaux,
in Analogue Gravity Phenomenology,
edited by D. Faccio, F. Belgiorno, S. Cacciatori,
V. Gorini, S. Liberati, and U. Moschella, Lecture Notes in
Physics Vol. 870 (Springer, Berlin, 2013), Chap. 6.


\bibitem{Doyle2013}	T.B. Doyle and J.F. McKenzie,
Quaestiones Mathematicae {\bf 36} (4), 487 (2013).



\end{thebibliography}
\end{document}